\documentclass[aps,prx,twocolumn,floatfix,superscriptaddress, nofootinbib]{revtex4-2}

\usepackage{amsmath}  
\usepackage{amsfonts} 
\usepackage{graphicx} 
\usepackage{epstopdf}
\usepackage{color}
\usepackage{siunitx}
\usepackage{comment}
\usepackage{commath}

\newcommand{\Poincare}{Poincar\'{e} }
\begin{document}

\title{Discovering the emergent nonlinear dynamics of acoustically levitated cube clusters}

\author{Annie Z. Xia}
\affiliation{Department of Physics, The University of Chicago, Chicago, Illinois 60637, USA}
\thanks{}

\author{Melody X. Lim}
\thanks{Current address: Department of Physics, Cornell University, Ithaca, New York 14853, USA}
\affiliation{Department of Physics, The University of Chicago, Chicago, Illinois 60637, USA}
\affiliation{James Franck Institute, The University of Chicago, Chicago, Illinois 60637, USA}

\author{Jason Z. Kim}
\thanks{Current address: Department of Physics, Cornell University, Ithaca, New York 14853, USA}
\affiliation{Department of Physics, Cornell University, Ithaca, New York 14853, USA}

\author{Bryan VanSaders}
\thanks{Current address: Department of Physics, Drexel University, Philadelphia, PA 19104, USA}
\affiliation{James Franck Institute, The University of Chicago, Chicago, Illinois 60637, USA}

\author{Heinrich M. Jaeger}
\affiliation{Department of Physics, The University of Chicago, Chicago, Illinois 60637, USA}
\affiliation{James Franck Institute, The University of Chicago, Chicago, Illinois 60637, USA}

\begin{abstract}

The complex behavior of many natural and engineered systems emerges from the interaction of a small number of effective degrees of freedom. Discovering the physical basis of the interactions between these degrees of freedom directly from experimental observations has been a longstanding challenge, particularly with respect to predicting the long-time dynamics of dynamical systems with unknown equations of motion. Here, we introduce a data-driven approach that is able to produce a generative model for the long-time dynamical behavior of systems with a weakly attracting manifold. We apply this method to an experimental dynamical system with two degrees of freedom: acoustically levitated pairs of cube-shaped particles, which cluster by sharing a single edge. In the acoustic trap, the center-of-mass of the cube cluster oscillates vertically about the levitation plane, while also oscillating about their flexible hinge-like connection. Depending on their initial condition, the hinge dynamics evolve about three distinct nonlinear dynamical attractors persisting for hundreds of cycles. In order to capture the underlying physics, we develop a numerical fitting procedure and extract a minimal nonlinear dynamical model that captures both the long-time dynamics of the cluster as well as the convergence onto the dynamical steady state. This dynamical model uncovers the nonlinear, non-reciprocal coupling between the center-of-mass motion and the hinge degree of freedom that stabilizes the dynamical attractors, which we subsequently confirm by independent finite-element methods. Our results demonstrate a novel data-driven method for the discovery of nonlinear models with long-timescale stable predictions. 

\end{abstract}

\maketitle

\section{Introduction}

Complex dynamics frequently emerges from nonlinear coupling between a small number of effective degrees of freedom in a wide variety of systems, including gene regulatory networks~\cite{ferrell2011modeling,karig2018stochastic,lederer2024statistical}, oscillating chemical reactions~\cite{tompkins2014testing}, ecological community dynamics~\cite{hu2022emergent}, and neuronal dynamics~\cite{hopfield1984neurons,abbott2005model,crunelli2010slow, colen_machine_2021}. In order to discover the governing equations underlying these dynamical systems, data is often modeled as a set of differential equations, with nonlinear coupling estimated by hand from physical first principles. Recent work in system identification and dynamical modeling has aimed to automatize these efforts by directly fitting nonlinear interaction terms to time-series data measured from complex systems
~\cite{bongard2007automated,schmidt2009distilling,wang2011predicting,daniels2015automated,brunton2016discovering,rudy2017data,schaeffer2017learning,schaeffer2017sparse,kaheman2020sindy,fasel2022ensemble,kaptanoglu2021pysindy,kim2025sigma}. A complementary set of neural-network-based tools have instead focused on accurate time-series prediction and generation without necessitating a set of pre-defined physical interactions~\cite{chen2018neural,pathak2018model,cuomo2022scientific,yu2025physics, yu_learning_2024, ghadami_data-driven_2022}. Both approaches have typically focused on predicting the short-time behavior of the system, i.e. accurate reconstructions of the local velocity and acceleration, or require strict conditions about the governing physics and functional form of the nonlinearities \cite{kaptanoglu2021promoting, colen_machine_2021, karniadakis_physics-informed_2021, bertalan_learning_2019}. However, dynamical systems which evolve about weakly attracting manifolds typically exhibit a large separation of scales between the short- and long-timescale behavior of the system. Such systems present a longstanding challenge to identifying interpretable and generative models, particularly in the presence of noise \cite{prokop2024biological, fasel_ensemble-sindy_2022}.

\begin{figure*}
    \centering
    \includegraphics[width=1.0\linewidth]{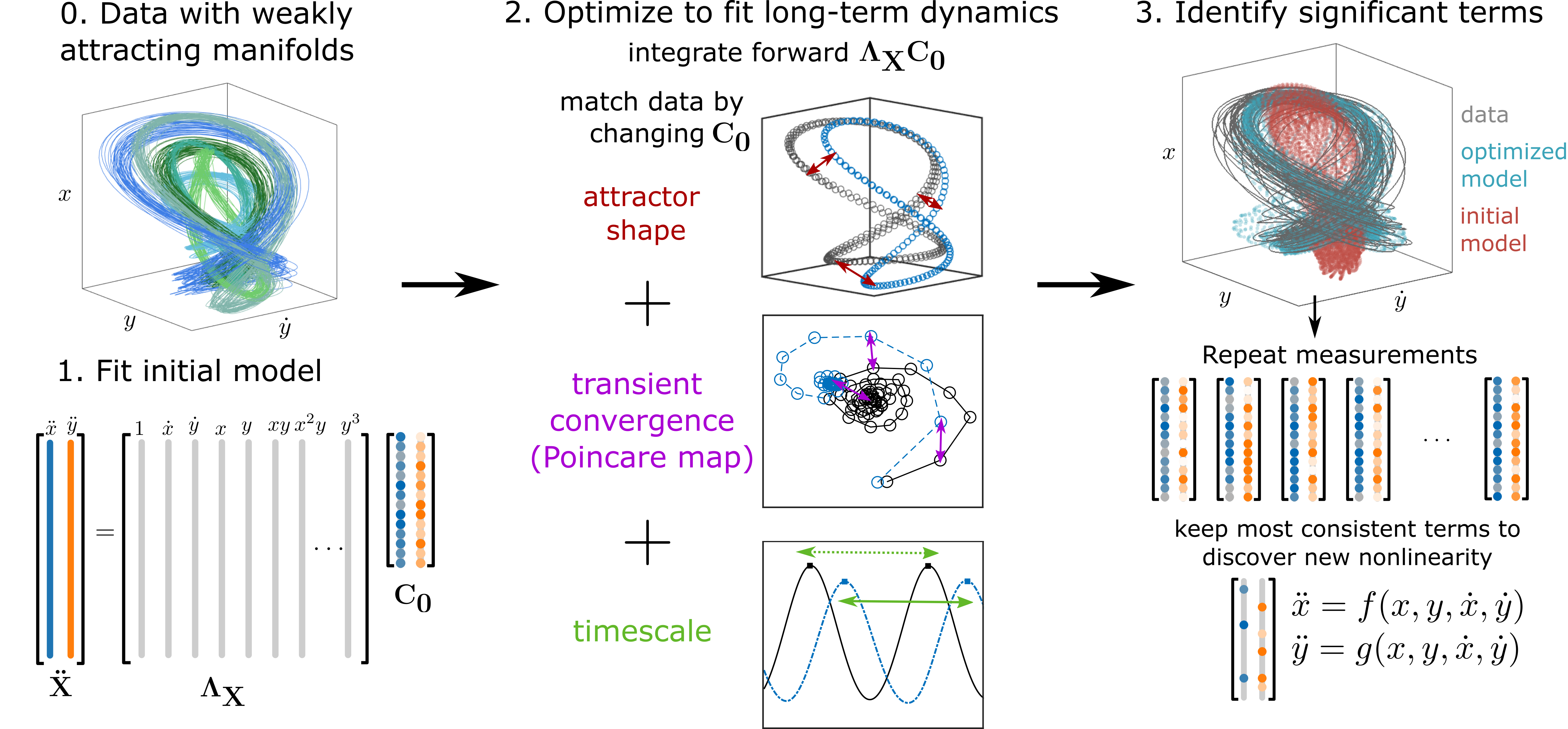}
    \caption{\textbf{Schematic of the fitting procedure.} We show sample data collected from an experimental system consisting of time-series data with two degrees of freedom $x$ and $y$. The instantaneous acceleration of the system is then reconstructed from linear and nonlinear combinations of the system variables by linear regression, resulting in an initial model $\mathbf{C_0}$ which fits instantaneous accelerations well but fails to capture long-time dynamical behavior. The dynamical equation is then refined by integrating the fitted dynamics forward, and modifying the coefficients to reconstruct long-time features of the original data. Finally, the coefficients are aggregated from repeat experiments, and only the most consistent terms are included in the final fit.}
    \label{fig:schematic}
\end{figure*}

Here, we present a novel data-driven approach (illustrated schematically in Fig.~\ref{fig:schematic}) to discover the governing equations underlying a dynamical system with two degrees of freedom. We begin this paper with a discussion of our data-driven approach for discovering governing dynamical equations in oscillatory systems with large separation in scales between short and long-time behaviors. We apply this method to an experimental system with two degrees of freedom: a pair of cube shaped particles levitated in an acoustic trap. Oscillations in the cluster bending angle couple nonlinearly to the center-of-mass motion of the cluster. We demonstrate that this nonlinearity manifests in the form of long-lived dynamical attractors in state space, and that convergence onto these dynamical attractors is robust for a wide range of initial conditions. In search of the physical basis of this nonlinearity, we develop a numerical fitting procedure that explicitly accounts for the long-term structure and stability of the attractor (schematically illustrated in Fig.~\ref{fig:schematic}). In order to produce robust estimates of the fitting parameters, we combine results from many different experimental datasets, retaining only the parameters that are statistically significant across all experiments. Using this fitting procedure, we develop a minimal model for the cluster dynamics that reveals a nonreciprocal nonlinear coupling between the cluster bending angle and the center-of-mass motion. Finite-element simulations confirm the same nonlinearity. Our results pave a way forward for robust parameter estimation and forecasting of weakly nonlinear, noisy physical systems. 

In order to accurately fit a dynamical model that produces stable long-timescale predictions, we develop a new approach where the attractor shape, stability, and convergence properties are directly incorporated into the cost function of the fitting procedure, shown schematically in Fig.~\ref{fig:schematic}. We apply this method on experimental observations of a dynamical system with two degrees of freedom which we call $x$ and $y$.

For the first step of our fitting procedure, we use existing methods for data-driven regression of the underlying dynamical equations~\cite{brunton2016discovering} to obtain reasonable initial coefficients that reconstruct the short-time experimental data. We assume dynamics containing polynomial terms up to third order in the state space $[x, y, \dot{x}, \dot{y}]$, and find the coefficients $\mathbf{C_0}$ which best reconstruct the instantaneous accelerations $[\ddot{x}, \ddot{y}]$. This fitting method is able to approximate the instantaneous accelerations reasonably well. However, integrating the dynamics forward in time produces attractor structures that do not match those observed in the experiment (red traces in step 3 of Fig.~\ref{fig:schematic}). In order to improve the predicted dynamics, we optimize the initial coefficients to match longer-time features in the data, including the steady state attractor shape, transient convergence trajectory, and the and the average oscillation timescale (step 2 in Fig.~\ref{fig:schematic}). We find that performing this optimization results in higher fidelity long-term reconstruction of the experimentally observed attractor manifolds. In order to find statistically significant nonlinearities, we then repeat this procedure for many experimental datasets, and compare the optimized coefficients. Identifying significant terms across the data ensemble, then re-optimizing with only the most significant coefficients leads to a minimal model, which is able to capture the observed long-term dynamical behavior with only a few key nonlinearities. 

\section{Methods}

\begin{figure}
\centering
\includegraphics[width = 1.0\columnwidth]{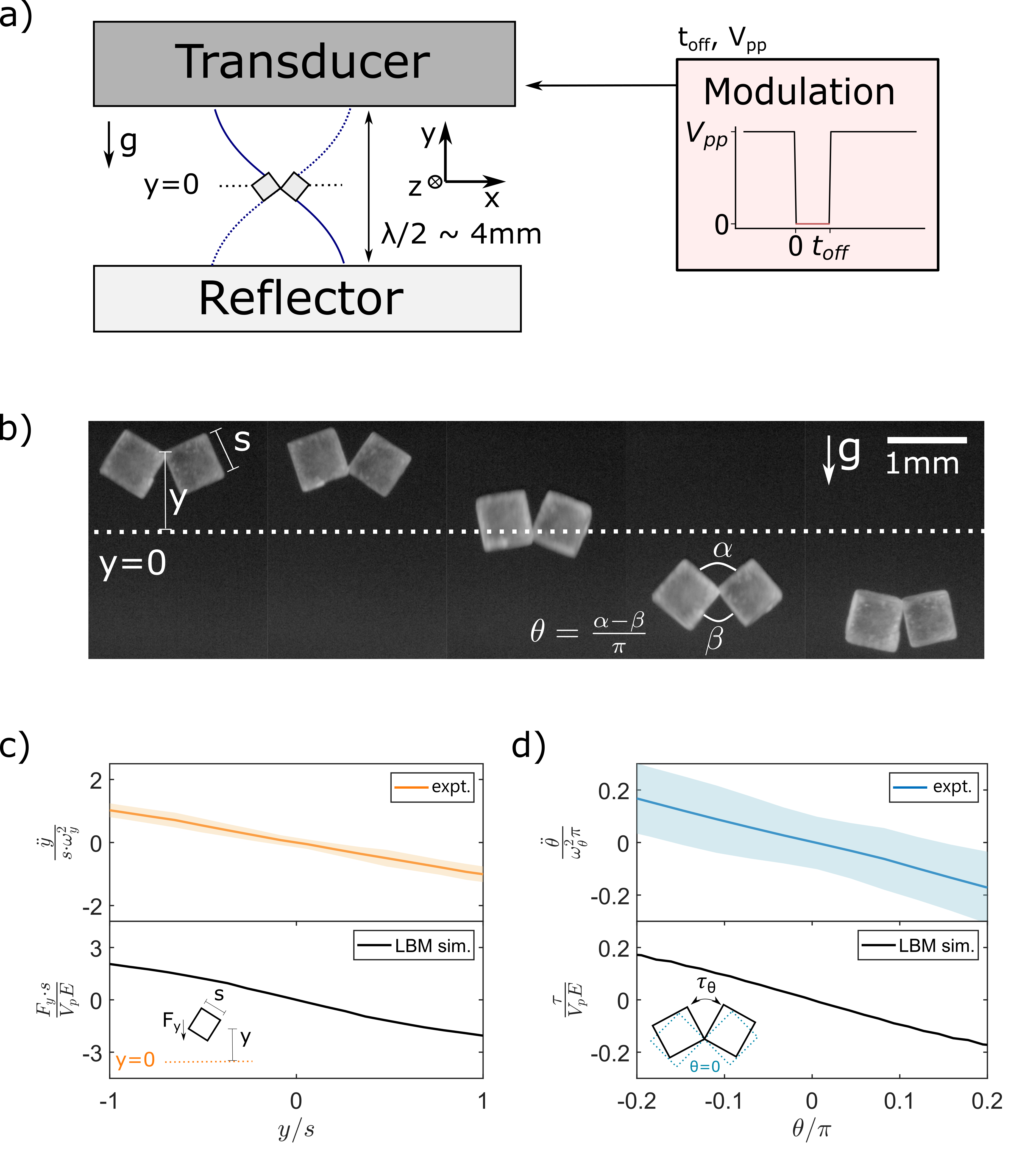}
\caption{\textbf{An acoustically levitated cube dimer is a granular hinge with an emergent elastic degree of freedom arising from particle geometry. }
(a) Diagram of experimental setup. The transducer emits ultrasound in air at $f_0 = 45.650$~kHz and wavelength $\lambda \approx 7.5$~mm. The reflector is placed at a resonant distance to generate a pressure standing wave with a single node. We apply a square wave modulation to turn off the field for some time $t_{\text{off}}$, injecting energy into the system. 
(b) Image of cube cluster dynamics over time, with subsequent images having a time difference of~$\Delta t = 4$~ms. The levitation plane defines $y=0$ (see Supplementary Information for details of the measurement), and we define $\theta$ as difference of the upper contact angle $\alpha$ and the lower contact angle $\beta$ divided by $\pi$.
(c) Restoring vertical force about the levitation plane due to the primary acoustic force. Experimental data (orange) is compared to the results of a Lattice-Boltzmann method (LBM) simulation (black) 
(d) Restoring torque about the contact point of the cluster. Experimental results (blue) are compared to the results of a LBM simulation (black).}
\label{fig:exp}
\end{figure}

We demonstrate the utility of our method by applying it to a model experimental system, consisting of a pair of sub-millimeter particles levitated in an acoustic trap. We establish an acoustic standing wave in air with a single node between a transducer and planar reflecting surface spaced one-half wavelength apart (Fig.~\ref{fig:exp}a, sound wavelength~$\lambda_s = 7.5$ mm). Computer control over the transducer allows for temporal modulation of the energy supplied to the acoustic cavity. Single particles levitate in the nodal plane, at a vertical position where the weight of the particle is balanced by the acoustic trapping force (which we define as~$y=0$). When a second particle is introduced to the acoustic trap, acoustic scattering due to the presence of the other particle generates attractive forces that cluster particles~\cite{lim_cluster_2019,lim_mechanical_2022,lim2024acoustic}. These attractive forces depend strongly on particle shape: pairs of rigid cubes (here grains of table salt with density $2160$ kg \si{\meter}$^{-3}$ and side lengths~$s$ between 400 to 700~$\mu$m) cluster acoustically by forming an edge-edge contact~\cite{lim_edges_2019}. The acoustic scattering force furthermore establishes an effective elasticity between the cubes in the cluster. In the underdamped environment of the acoustic trap, this elasticity results in oscillatory bending motion about the point of contact. We define $\theta=\frac{\alpha-\beta}{\pi}$ to be the normalized difference between the top angle $\alpha$ and the bottom angle $\beta$. $\theta$ takes values between $\pm1$, which correspond to a cube dimer contacting face-to-face. This angle will oscillate around $\theta=0$, corresponding to a corner-to-corner contact where the top angle and bottom angle are both $\pi/4$. At the same time, due to restoring forces from the acoustic trap, the cluster undergoes center-of-mass oscillations about its equilibrium levitation height (Fig.~\ref{fig:exp}b).  

We characterize the elasticity of each of these degrees of freedom by measuring the displacement of the cube cluster from its equilibrium position. We estimate the restoring elastic force on the vertical center-of-mass~$y$ of the cube cluster by measuring the vertical acceleration~$\ddot{y}$. Plotting~$\ddot{y}$ as a function of~$y$ reveals that for vertical displacements smaller than the side length of a cube~$y<s$, the restoring force is approximately linear in~$y$, giving rise to simple harmonic motion about the levitation plane. These observations are in good qualitative agreement with the results of Lattice-Boltzmann Method (LBM) simulations (Fig.~\ref{fig:exp}c). Similarly, plotting the bending angle acceleration~$\ddot{\theta}$ as a function of the bending angle~$\theta$ reveals a linear restoring torque for small bending angles, again in good agreement with the results of LBM simulations (Fig.~\ref{fig:exp}d). 

\begin{figure*}
\centering
\includegraphics[width = 2\columnwidth]{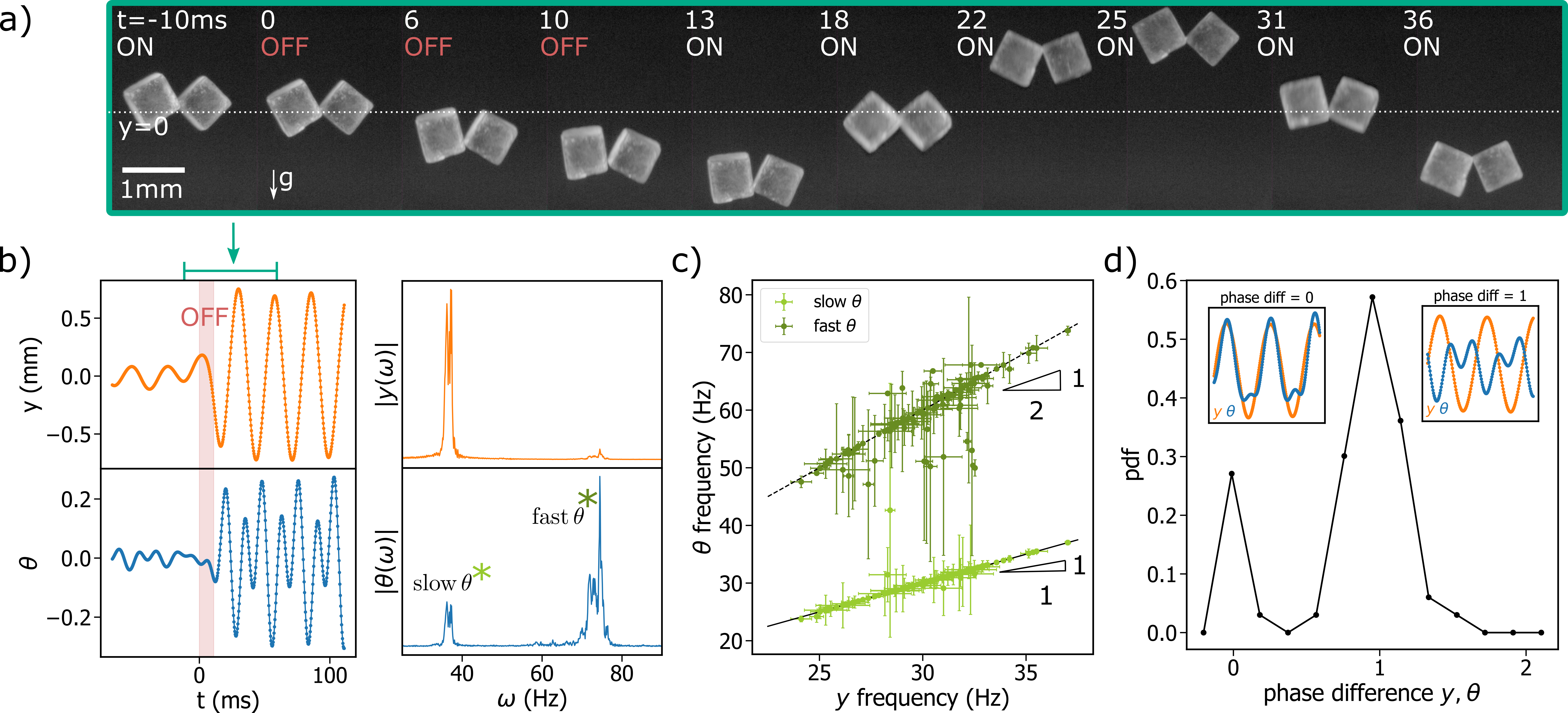}
\caption{\textbf{External modulation controllably excites both oscillatory hinge modes.}
(a) Image sequence showing field modulation. Prior to modulation ($t<0$), the cluster is stably levitated in the acoustic field. At $t=0$ the field is turned off and the cubes fall under gravity. At $t=13$ms, the field is turned on again and the clusters experience restoring forces, giving rise to oscillations in both degrees of freedom.
(b) (Left) Smoothed traces of $y$ and $\theta$ between the cubes before, during, and after the modulation. (Right) Discrete Fourier transform of the $y$ and $\theta$ signals.
(c) Ratio between $\omega_y$ and the two slow and fast angle components, plotted in light green and dark green respectively. (d) Plot of the probability density function of the phase difference between the~$y$ and~$\theta$ oscillations (examples for phase difference of 0 and 1 plotted in insets). 
}
\label{fig:mod}
\end{figure*}

In order to access dynamics beyond the linear regime, we briefly turn off the acoustic field, allowing the center-of-mass of the cube cluster to accelerate downwards under the effect of gravity (series of still images shown in Fig.~\ref{fig:mod}a, Supplemental Movie 1). Since the removal of the acoustic field also removes the restoring torque keeping~$\theta=0$, the cubes are also free to pivot about their shared edge. Once the acoustic field is turned back on, the cube cluster experiences forces and internal torques according to its new configuration, which are now far from the mechanical equilibrium conditions at~$(y,\theta)=0$. As a result, this brief field modulation drives large amplitude oscillations in both~$y$ and~$\theta$. 

To quantify these observations, we plot the resulting oscillations in~$y$ and~$\theta$ as a function of time before and after the modulation (Fig.~\ref{fig:mod}b), together with their discrete Fourier transform. We find that after field modulation, the center-of-mass oscillations appear sinusoidal, with a single frequency component, consistent with the expectation that the cluster behaves as a simple harmonic oscillator with linear restoring force (Fig.~\ref{fig:exp}c). In contrast, oscillations in the bending angle consist of two separate frequency components. Plotting the two frequency components as functions of the y-frequency over many experiments (Fig.~\ref{fig:mod}c) reveals that the slow frequency component is locked to the y-frequency, while the fast frequency is double the y-frequency. Furthermore, this frequency doubling is accompanied by phase-locking between the y-oscillations and the slow~$\theta$ frequency oscillations. Plotting the probability density function of the phase differences between oscillations in~$y$ and~$\theta$ reveals that the distribution of phase differences is bimodal, with peaks where~$y$ and~$\theta$ are either exactly in phase or antiphase (Fig.~\ref{fig:mod}d). This frequency and phase locking suggests the existence of a nonlinear coupling between cluster shape and its vertical displacement in the acoustic trap. 

\section{Results}

To further characterize this nonlinear coupling, we turn to the dynamical phase-space of the cube cluster. Since the system dynamics are underdamped, the dynamic trajectory of a cube dimer is characterized by 4 state variables: $\{y, \dot{y}, \theta, \dot{\theta} \}$. As~$y$ is close to harmonic, we choose to visualize on the 3-dimensional axes $\{\theta, \dot{\theta},y\}$. Plotting our data on these axes reveals the existence of three distinct attracting trajectories in phase space (examples plotted as blue, red, and green traces in Fig.~\ref{fig:classes}a, see Supplementary Movies 2-4 for dynamics). These trajectories persist for over one hundred cycles in~$\theta$ without changing in amplitude. Qualitatively, the key differences between the three trajectories are in the amplitude of the blue class compared to the red and green class, while the red and green trajectories differ in the relative phase between $y$ and $\theta$ oscillations (appearing as a relative rotation in the three-dimensional state space, and corresponding to the two classes observed in Fig.~\ref{fig:mod}d). 

\begin{figure*}
\centering
\includegraphics[width = 2\columnwidth]{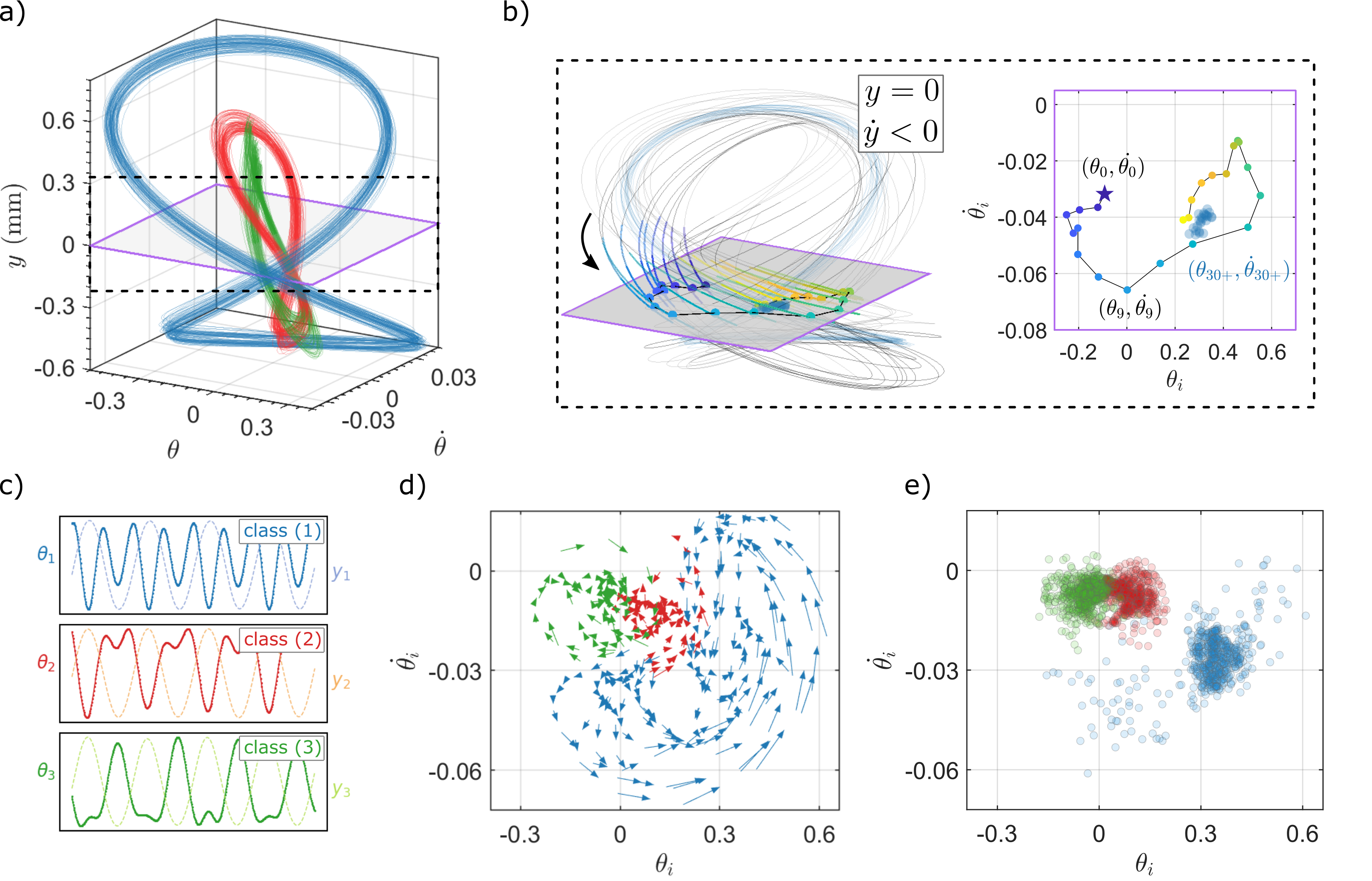}
\caption{\textbf{Phase locking in oscillations leads to three distinct dynamic memories persisting over hundreds of cycles}
(a) Sample trajectories for three distinct attractor types (100 cycles of each), projected onto the~$\theta$,$\dot{\theta}$,$y$ axes. The $y=0$ hyperplane is shown in purple.
(b) Example of the \Poincare section, together with resulting \Poincare map.  We define the \Poincare section as the $y=0$ hyperplane (purple), and take the \Poincare map to be the point in each cycle where the data first crosses the section in the $\dot{y}<0$ direction (from above in this projection). Purple box: resulting \Poincare map for 60 cycles of data (color corresponding to time after drop for the first 30 cycles. Cycles 30-60 are plotted in light blue). 
(c) Sample trajectories of~$\theta(t)$ (dark) and~$y(t)$ (light) for the three classes shown in (a). 
(d) Sample flows on the \Poincare maps for the first $10$ ($20$) points in each trajectory, separated by attractor class. $10$ trajectories from each class are included. A Savitzky-Golay filter of order $2$ and window length $5$ is applied to the \Poincare map to smooth the flows. (e) Aggregated \Poincare maps for the last $60$ points in each trajectory, colored by attractor class. $10$ trajectories from each class are included.}
\label{fig:classes}
\end{figure*}

In order to confirm the general dynamical stability of these attractors across the experimental dataset, we define a \Poincare section to further reduce the data dimensionality. This \Poincare section, and accompanying \Poincare map allow us to quantify whether the attractor location is generally fixed across the observed experimental data, and additionally allows for quantification of the convergence of off-manifold points onto the attractor manifold. We define the \Poincare section as the hyperplane $y=0$, and the corresponding \Poincare map such that the $i$-th point of the \Poincare map $(\theta_i, \dot{\theta_i})$ is the $i$-the crossing of $y=0$ from above as the system evolves in time (example shown in Fig.~\ref{fig:classes}b). We find that the \Poincare map consists of two distinct parts: first, evolution from the initial condition along some transient trajectory, and second, a steady state fixed point where the map returns to the same location in the \Poincare section for the rest of the observed cycles. Aggregating experimental data reveals that both the transient convergence trajectory (Fig.~\ref{fig:classes}d) and the fixed point location (Fig.~\ref{fig:classes}e) are consistent across multiple observations. Furthermore, each of the three attractors (classified by phase difference and amplitude, steady-state time-series shown in Fig.~\ref{fig:classes}c) is distinguished by a distinct convergence path, together with its unique steady state parameterized by $(\theta_i, \dot{\theta_i})$. In a system with only two degrees of freedom and four state variables, the coexistence of these three attractors points to the existence of an emergent nonlinearity that couples the cluster shape and its vertical position in the acoustic trap. 

\begin{figure*}
\centering
\includegraphics[width = 2.0\columnwidth]{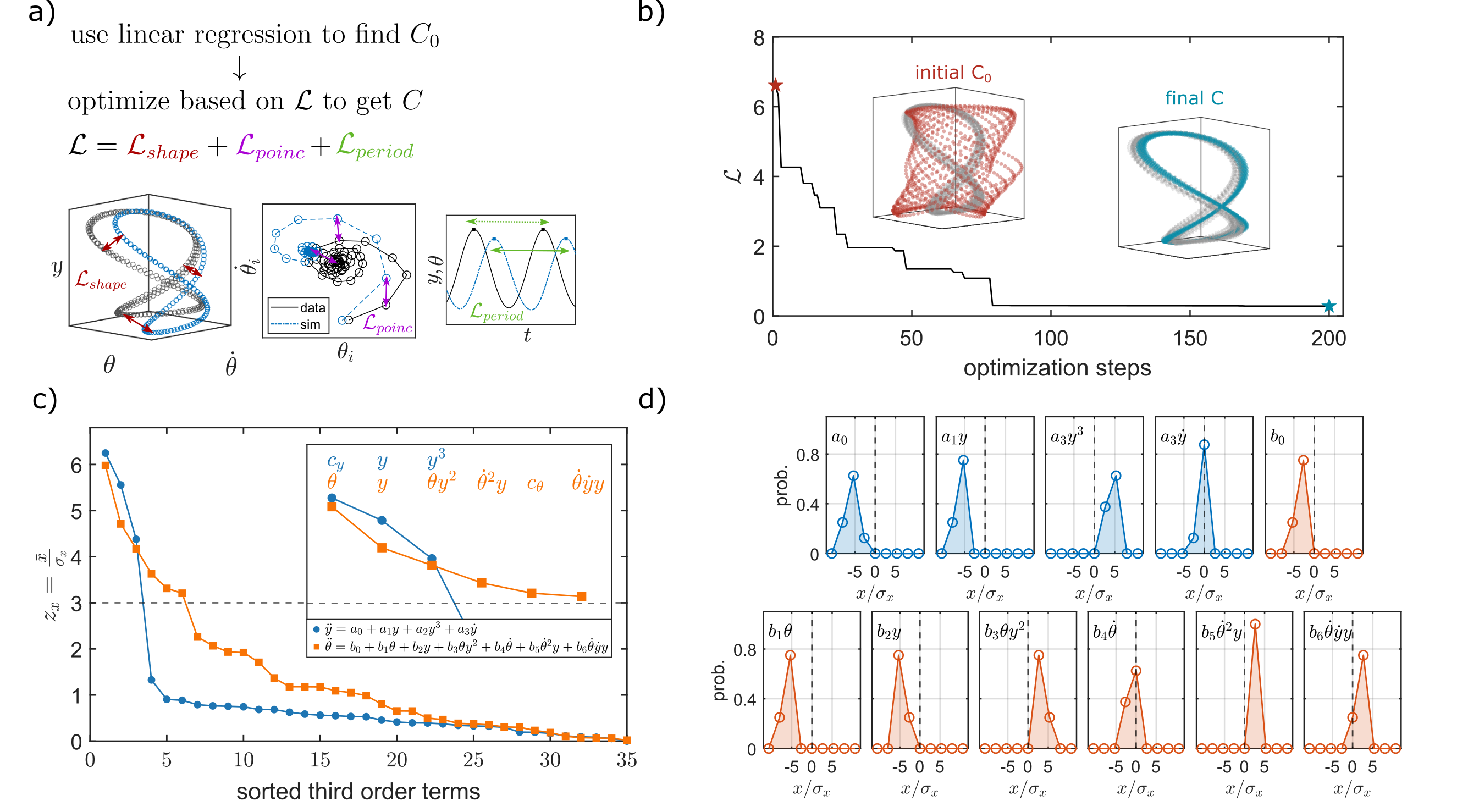}
\caption{\textbf{Discovering the governing nonlinearity via model fitting.} (a) Schematic of our parameter fitting procedure. We first find the best linearly fitting coefficients $\mathbf{C_0}$ which minimizes the error $\norm{\mathbf{\ddot{X}} - \mathbf{\Lambda}_X \mathbf{C_0}}$ between experimental acceleration $\mathbf{\ddot{X}}$ and polynomial combinations of our state variables $\mathbf{\Lambda}_X$. These coefficients $\mathbf{C_0}$ are then used as initial conditions for gradient descent over the loss function $\mathcal{L}$ which has three terms, each penalizing different deviations of simulated, integrated dynamics from the data time-series. 
(b) Example optimization trajectory starting from initial $\mathbf{C_0}$ and reaching final $\mathbf{C}$.
(c) Z-scores for second order fits of $\ddot{y}$ and $\ddot{\theta}$ after optimization over 10 large attractor (class 1) trajectories. Discarding terms with a z-score lower than $3$, and including physical terms for viscous damping, we attain two nonlinear equations for $\ddot{y}$ and $\ddot{\theta}$.
(d) Normalized coefficient distributions for the eleven significant terms used in the reduced model.}
\label{fig:gd}
\end{figure*}

Our experimental observations, consisting in changes over time of two variables, are naturally represented by a pair of differential equations, with an unknown nonlinear coupling between the two variables~$\theta$ and~$y$. In order to discover the governing equations, and particularly the form of the nonlinearity that is responsible for the observed attractor structure, we turn to data-driven system identification methods. Previous work on system identification methods is generally focused on accurate short-term trajectory prediction, i.e. instantaneous accelerations and velocities. Our data presents a unique challenge in that the nonlinearity stabilizes relatively long-timescale structures. In order to accurately fit a dynamical equation that produces stable long-timescale predictions, we apply the new approach described previously, where the attractor shape, stability, and convergence properties are directly incorporated into the cost function of the fitting procedure. 

For the first step of our model fitting procedure, we use existing methods for data-driven regression of the underlying dynamical equations~\cite{brunton2016discovering} to obtain reasonable initial coefficients that reconstruct the short-time experimental data.  Specifically, for a single trial, we construct an array~$\mathbf{X}$ containing time-series of our four state variables:~$\mathbf{X} = [y, \dot{y}, \theta, \dot{\theta}]$. Our objective is to reconstruct the observed accelerations $\mathbf{\ddot{X}} = [\ddot{y}$, $\ddot{\theta}]$ by linear and nonlinear combinations of the four state variables. In order to do this without making any assumptions about the form of the nonlinearity, we create a nonlinear library of polynomial functions up to third order, which we call $\mathbf{\Lambda}_X$.  There are thirty five third order terms (for example, $y,\theta^3,\dot{\theta}y^2$, etc.). In order to avoid spurious terms in this dynamical fitting, we choose the equilibrium levitation height~$y_0$ in the rest frame of the cluster (See Appendix D for further discussion on $y_0$). The task of finding the governing equation that best explains the instantaneous (short-term) dynamics can thus be framed as solving the linear regression problem~$\mathbf{\ddot{X}} = \mathbf{\Lambda}_X \mathbf{C}$, where~$\mathbf{C}$ is the matrix of coefficients that determines the contribution of each term in $\mathbf{\Lambda}_X$ to the instantaneous acceleration. 

Although this fitting method is able to reasonably approximate the instantaneous acceleration given the current state of the dynamical system, initial conditions integrated forward with these dynamics deviate significantly from experimental trajectories (example shown in Fig.~\ref{fig:gd}b, see Appendix C for further characterization). As a result, the initial linear regression method produces coefficients that are unable to generate the same long-time dynamics as observed in the experiment. 

\begin{figure*}
\centering
\includegraphics[width = 2\columnwidth]{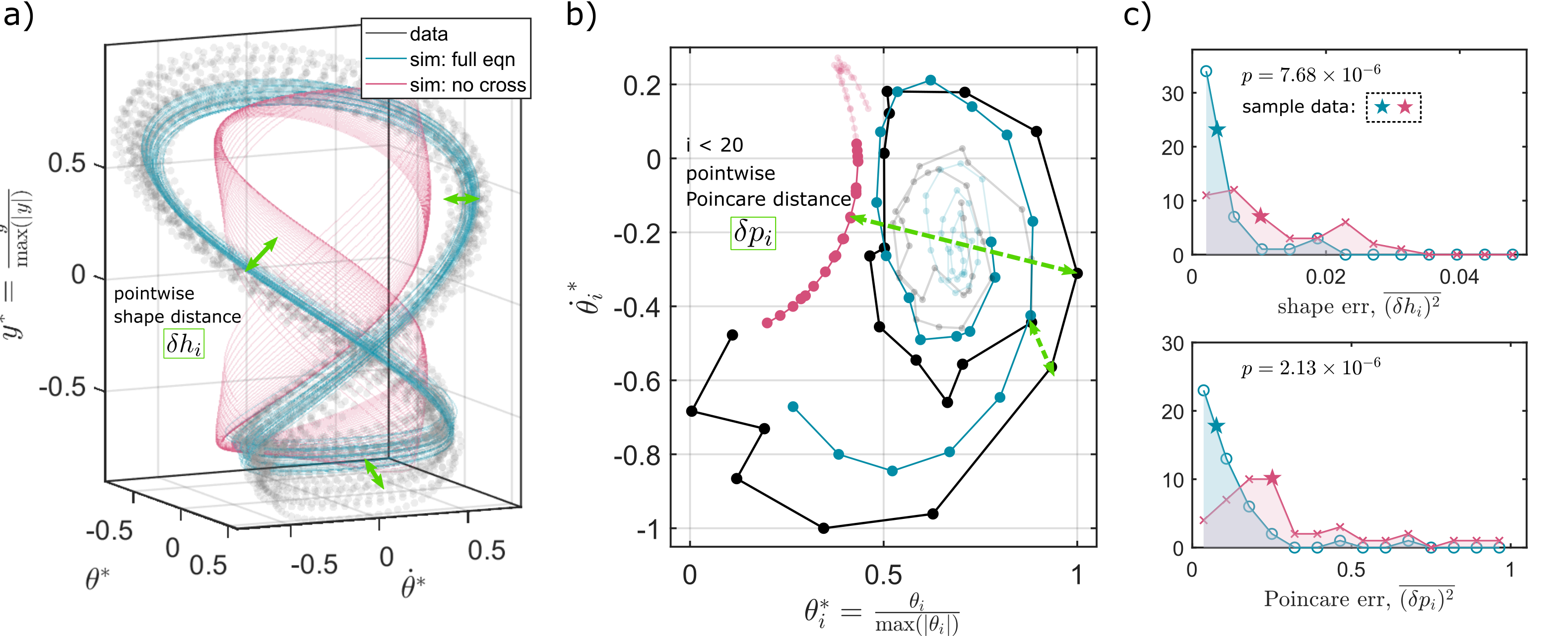}
\caption{\textbf{Simple nonlinear model captures shape and convergence of dynamical memories} (a) Sample attractor trajectory with (blue) and without (pink) cross coupling term $b_4 \theta y$. Green lines represent $\delta h_i$, the minimum distance between a particular simulated point and the data trajectory. State variables are all normalized according to $x^* = \frac{x}{\max(|x|)}$. The pointwise squared error $\delta h_i^2$ averaged over all points gives $\overline{\delta h_i^2}$. 
(b) Sample \Poincare section with (blue) and without (pink) cross coupling term $b_4 \theta y$. Green lines represent $\delta p_i$, the normalized distance between the simulation and data. The state variables are normalized according to $x^* = \frac{x}{\max(|x|)}$
(c) Distribution of $\overline{\delta h_i^2}$ for 55 datasets. The stars show values of $\overline{\delta h_i^2}$ from the sample trajectories in (a) and (b). Reported p-values are from a signed rank test with null hypothesis $H_0: \mu_{f} - \mu_{o} = 0$. (d) Distribution of $\overline{\delta p_i^2}$ for 55 datasets.}
\label{fig:simulation}
\end{figure*}

To refine the dynamics and correctly capture the attractor structure observed in experiments, the second step of our fitting procedure performs gradient-free optimization on the coefficients obtained via linear regression. This optimization explicitly acts to minimize a loss function $\mathcal{L}$ that contains information about the long-term structure and stability of the experimental data. In order to do so, we construct $\mathcal{L}$ as the sum of three terms which respectively penalize deviations from the shape of the experimentally measured steady-state dynamical attractor, the convergence properties (as measured on the \Poincare map), and oscillation frequency (schematically depicted in Fig.~\ref{fig:gd}a). Performing an optimization to minimize~$\mathcal{L}$ by changing~$\mathbf{C}$ produces long-time dynamics that match the attractor structure in the experiments more closely than those obtained via linear regression (Fig.~\ref{fig:gd}b, See Appendix C for details about model optimization). 

In order to make this model estimation procedure robust against noise in the experimental system and correctly generalize across multiple datasets, we eliminate terms with low statistical significance from the regression. Specifically, we examine the distribution of coefficient values across repeated experimental trials for the large amplitude attractor (class 1), and plot the z-score (mean divided by standard deviation) for the values of~$\mathbf{C}$ in Fig.~\ref{fig:gd}(c) for both $\ddot{y}$ and $\ddot{\theta}$. In doing so, we find that the vast majority of the third-order terms have small z-scores, suggesting that the data can be well explained by a sparse subset of the terms. We identify these statistically significant terms by applying a cutoff of $z=3$, and adding viscous damping terms, to obtain two simplified dynamical equations:

\begin{equation}
\ddot{y} = a_0+ a_1y + a_2 y^2 + a_3\dot{y} 
\label{eq:y1}
\end{equation}

\begin{equation}
\ddot{\theta} =b_0+ b_1 \theta + b_2 y + b_3 \theta y^2 + b_4 \dot{\theta} +b_5 \dot{\theta^2}y +b_6 \dot{\theta}\dot{y}y \, .
\label{eq:th1}
\end{equation}

Plotting the distribution of ~$a_0$ to~$a_3$, and~$b_0$ to~$b_6$ across experimental trials (Fig.~\ref{fig:gd}d) reveals roughly Gaussian distributions of the coefficients, with non-zero mean. Our results show that the observed nonlinearities of the experimental data are well explained by a sparse set of three nonlinear terms that couple the bending angle of the cube cluster to its center-of-mass position (with coefficients $b_4$, $b_5$ and $b_6$). These nonlinear terms are robustly observed across multiple experimental trials, suggesting that the consistently observed nonlinear dynamics (Fig.~\ref{fig:classes}) stem from a single coherent governing equation. Note that the constant terms $a_0$ and $b_0$ in the minimal model can be estimated from the other parameters and the choice of rest frame $y_0$ and do not need to be optimized separately (See Appendix D for details)

In order to independently validate the numerically discovered nonlinearity, we performed finite-element simulations (see Supplementary Information). Due to computational cost, these finite-element simulations are able to capture only static forces and torques (i.e. no nonlinearities proportional to~$\dot{y}$ or~$\dot{\theta}$). Nevertheless, these lowest-order simulations also show a contribution to the restoring torque of the form~$\ddot{\theta}\propto \theta y^2$, corresponding to the term~$b_4$ above, and independently validating the numerically discovered nonlinearity to lowest order. 

We now examine whether this simplified governing equation, with the three nonlinear terms, is necessary and sufficient to capture the nonlinear dynamics observed in the experiment. We do so by integrating the simplified equation of motion forward from the initial conditions of the experiment, and observing the long-time behavior, as well as the convergence onto the attractor manifold. To assess the role of the nonlinear terms in the system dynamics, we perform the forward integration both for the equation of motion Eq.~\ref{eq:th1}, as well as a version where we omit the nonlinear cross terms (i.e. set~$b_4,b_5,b_6=0$ and redo the optimization). 

Plotting the phase-space trajectory of the sample forward-simulated data reveals that the full equation of motion closely conforms to the experimentally observed attractor structure (blue simulated data vs gray data points in Fig.~\ref{fig:simulation}a), while omitting the nonlinear term causes deviations from the shape of the observed attractor structure (pink simulated data vs gray data points in Fig.~\ref{fig:simulation}a). Examining the \Poincare map for the first few iterations reveals that convergence onto the attractor manifold also takes a significantly different trajectory for the case without nonlinearity (Fig.~\ref{fig:simulation}b). 

Quantitatively, we evaluate the discrepancy between the forward-simulated equations of motion and the experimental results using the shape distance~$\delta h_i$ and the distance between the \Poincare maps~$\delta p_i$ (see Appendix C for further details, green arrows in Fig.~\ref{fig:simulation}a,b for schematic representation).  Evaluating the average shape error $\bar{(\delta h)^2}$ and average convergence error $\bar{(\delta p)^2}$ reveals that the full equations of motion have small shape error and convergence error (blue data in Fig.~\ref{fig:simulation}c,d). In contrast, omitting the nonlinear coupling significantly increases both the shape and convergence error (pink data in Fig.~\ref{fig:simulation}c,d).  For both the shape error and \Poincare convergence error, a signed rank test comparing the error distributions of the full dynamics and a dynamics with no cross term has a p-value less than $10^{-5}$. Our results signify that emergent nonlinearity in the form of the $\theta y^2$ cross term is essential in correctly capturing the co-existence, shape, and convergence of the attractor structure seen in experiment.

\section{Discussion}

We have shown that emergent acoustic-shape interactions give rise to rich dynamical behaviors. Here, in an acoustically levitated system with only one shape degree of freedom, an emergent nonlinear restoring torque stabilizes three distinct attracting trajectories. The steady state oscillations of these attractor-like trajectories are long lived and can exhibit convergence times an order of magnitude larger than the typical timescales of the system. We are able to capture this long time behavior in a minimal dynamical model by minimizing a loss function which captures attractor convergence and steady state shape. Through this dynamical fitting and finite element simulations, we find a nonlinear cross-coupling term $\theta y^2$ which is necessary to recover the coexistence, convergence, and steady state structure of these attractors. 



We anticipate that our data-driven model-discovery procedure will be particularly useful for noisy experimental data where the attracting manifolds are only weakly nonlinear. In these cases, applying existing sparse regression methods may not correctly identify the driving nonlinearities, as weakly nonlinear terms may only contribute relatively little to the local system accelerations. Furthermore, fitting procedures that rely exclusively on matching instantaneous derivatives may not be robust to sources of experimental noise that affect the time derivatives, for instance if there are small variations in the oscillation period over time. Our method reduces reliance on these short-time variations and boosts the significance of weakly nonlinear terms by including the \Poincare map as a coarse-grained representation of the system (evaluated at every oscillation period $T$, rather than every time point $t$). Combining this approach with ensemble-based measures of statistical significance allows for accurate dynamical fitting together with an interpretable minimal model. We envision applications and extensions of our method to a wide variety of other physical and biological dynamical systems to obtain interpretable and generative minimal models from noisy experimental data.

\section*{Acknowledgments}

A.Z.X. and B.VS. acknowledge support from the MRSEC program of the NSF under DMR-2011854. M.X.L. and H.M.J. acknowledge support from NSF DMR-2104733. J.Z.K. was supported by postdoctoral fellowships from Bethe/KIC/Wilkins, Mong Neurotech, and the Eric and Wendy Schmidt AI in Science program of Schmidt Sciences, LLC.


%

\end{document}